\documentclass[%
reprint,
superscriptaddress,
preprintnumbers,
nofootinbib,
amsmath,amssymb,
aps,
prl,
floatfix,
]{revtex4-2}

\usepackage{enumitem}
\usepackage{hyperref}
\usepackage{graphicx}
\usepackage{dcolumn}
\usepackage{bm}

\usepackage{epsfig}
\usepackage{amsmath}
\input{epsf}
\usepackage{psfrag}
\usepackage{xcolor}
\usepackage{pdfpages}
\usepackage{subcaption}
\usepackage{booktabs}
\usepackage[normalem]{ulem}
\usepackage{siunitx}
\usepackage{listings} %
\makeatletter
\AtBeginDocument{\let\LS@rot\@undefined}
\makeatother
\newcommand{\bea}{\begin{eqnarray}}
\newcommand{\eea}{\end{eqnarray}}

\newcommand{\nn}{\nonumber}

\newcommand{\ket}[1]{\left| #1\right>}

\definecolor{myBlack}{rgb}{0.45,0.3,0.00}

\begin{document}

\preprint{CPTNP-2025-013}

\title{
Scalable Quantum State Preparation via  Large-Language-Model-Driven Discovery
}
 
\author{Qing-Hong Cao}
\email{qinghongcao@pku.edu.cn}
 \affiliation{School of Physics, Peking University, Beijing 100871, China}
 \affiliation{Center for High Energy Physics, Peking University, Beijing 100871, China}

 \author{Zong-Yue Hou}
\email{zongyuehou@stu.pku.edu.cn}
 \affiliation{School of Physics, Peking University, Beijing 100871, China}

 \author{Ying-Ying Li}
 \email{liyingying@ihep.ac.cn}
 \affiliation{Institute of High Energy Physics, Chinese Academy of Sciences, Beijing 100049, China}

 \author{Xiaohui Liu}
 \email{xiliu@bnu.edu.cn}
 \affiliation{Center of Advanced Quantum Studies, School of Physics and Astronomy, Beijing Normal University, Beijing, 100875, China}
 \affiliation{Key Laboratory of Multi-scale Spin Physics, Ministry of Education, Beijing Normal University, Beijing 100875, China}

 \author{Zhuo-Yang Song}
 \email{zhuoyangsong@stu.pku.edu.cn}
 \affiliation{School of Physics, Peking University, Beijing 100871, China}

 \author{Liang-Qi Zhang}
 \email{liangqizhang@pku.edu.cn}
 \affiliation{School of Physics, Peking University, Beijing 100871, China}

 \author{Shutao Zhang}
 \email{shutaozhang@pku.edu.cn}
 \affiliation{School of Physics, Peking University, Beijing 100871, China}

 \author{Ke Zhao}
 \email{ke-zhao@pku.edu.cn}
 \affiliation{School of Physics, Peking University, Beijing 100871, China}

\begin{abstract}
Efficient quantum state preparation remains a central challenge in first-principles quantum simulations of dynamics in quantum field theories, where the Hilbert space is intrinsically infinite-dimensional. Here, we introduce a large language model (LLM)-assisted framework for quantum-circuit design that systematically scales state-preparation circuits to large lattice volumes. Applied to a $1$+$1$d $XY$ spin chain, the LLM autonomously discovers a compact 4-parameter circuit that captures boundary-induced symmetry breaking with sub-percent energy deviation, enabling successful validation on the \texttt{Zuchongzhi} quantum processor. Guided by this insight, we extend the framework to 2+1d quantum field theories, where scalable variational ansätze have remained elusive. For a scalar field theory, the search yields a symmetry-preserving, 3-parameter shallow-depth ansatz whose optimized parameters converge to size-independent constants for lattices $n \ge 4$, providing, to our knowledge, the first scalable ansatz for this class of 2+1d models. Our results establish a practical route toward AI-assisted, human-guided discovery in quantum simulation.
\end{abstract}

\maketitle

\textbf{\textit{  Introduction.}} 
Quantum computing promises exponential advantages in solving certain classes of complex computational problems across diverse fields, ranging from combinatorial optimization~\cite{Harrigan:2020dvo, Ebadi:2022oxd, Zhu:2024yhe} and quantum chemistry~\cite{PhysRevA.90.022305} to fundamental physics. Recent advances in quantum hardware and algorithmic design have driven rapid progress in exploring its applications within various subfields of physics. In particular, significant efforts have been devoted to establishing theoretical frameworks for simulating nonperturbative dynamics in quantum field theories—problems that lie beyond the reach of classical computational methods~\cite{Feynman:1981tf, Jordan:2012xnu, Bauer:2022hpo, DiMeglio:2023nsa, np-review, Fang:2024ple}. While these theoretical developments have laid crucial groundwork, translating them into concrete computational implementations on noisy intermediate-scale quantum (NISQ)~\cite{Preskill:2018jim} hardware remains a formidable challenge. Among the key bottlenecks is the efficient preparation of physically relevant initial states, a prerequisite for realizing meaningful quantum simulations.

As a leading strategy for state preparation on NISQ devices, Variational Quantum Algorithms (VQAs)~\cite{McClean:2015vup, Farhi:2014ych, Cerezo:2020jpv} have become the default approach. However, conventional VQAs typically rely on manually designed ansatz circuits, whose construction demands substantial expert intuition and often lacks scalability to larger or more complex systems such as those encountered in quantum field simulations. Moreover, these ansätze frequently require deep circuits with numerous variational parameters, which exacerbate optimization difficulties and can lead to barren plateau phenomena~\cite{McClean:2018jps, Martyniuk:2024bzu}.

In this work, we tackle these physical bottlenecks by introducing a new methodology for quantum circuit design in which Large Language Model (LLM)-driven search and human physical intuition reinforce each other to uncover efficient, interpretable, and hardware-ready quantum circuits.  
We demonstrate this paradigm with two key findings that highlight its collaborative components.
First, we show the LLM acting as an {\it intuition catalyst}: when provided with only minimal templates, it autonomously discovered the need for a spatially modulated ansatz to capture the ground state of the open-boundary $XY$ model. This machine-generated insight enabled us (the human experts) to collaboratively refine it into the final compact, high-fidelity circuit.

Building on the success in $1$+$1$d theories, we show how this paradigm addresses a key open problem: the construction of a demonstrably scalable variational quantum ansatz for 2+1d quantum field theories, a primary near-term goal for quantum devices. We focus on a self-interacting scalar field theory, which illustrates essential issues for eventually simulating more complex gauge theories such as QED and QCD~\cite{Jordan:2012xnu}. Inspired by the emergent ``intuition" from the $1$+$1$d theory, we explicitly guided the LLM with physical knowledge (i.e., symmetries) to tackle the much harder $2$+$1$d scalar field theory.

This synergy, combining the LLM's novel search capabilities with targeted human guidance, proved critical. It generates efficient, scalable ansätze for large-lattice ground states previously intractable for VQAs, successfully navigating the optimization landscape. The resulting shallow, low-parameter ansätze are inherently more resilient to barren plateaus. The framework is now open source at \href{https://github.com/IdeaSearch/IdeaSearch-QC}{Github}.

\begin{figure}
   \begin{center}
    \includegraphics[scale=0.25]{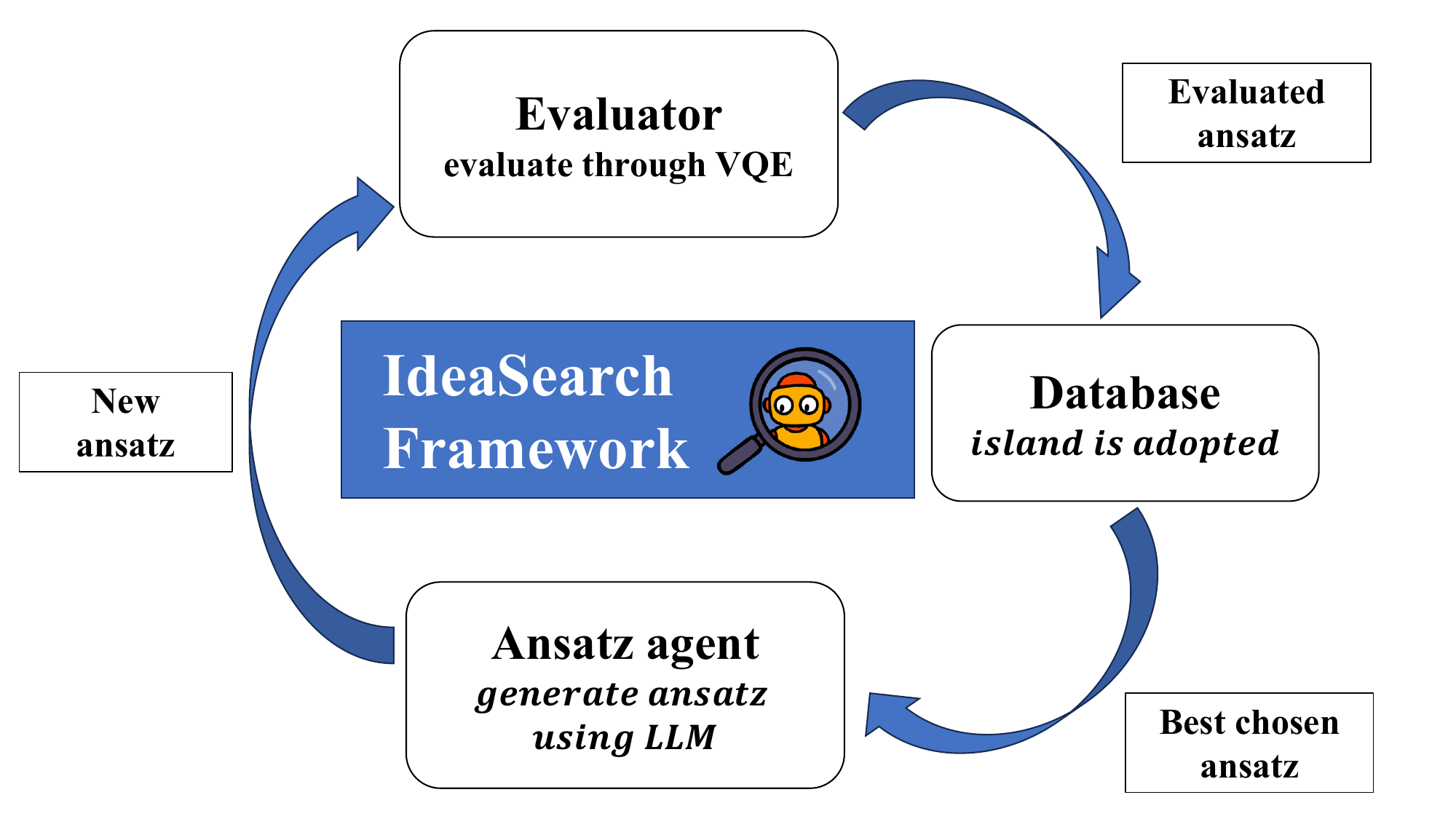} 
 \caption{Integrated workflow for quantum ansatz design.  
 }
   \label{fg:workflow-llm-qc}
  \end{center}
   \vspace{-5.ex}
 \end{figure}

\textbf{\textit{ The Generative Agent Framework. }} The framework, IdeaSearch~\cite{ideasearch2025}, is an iterative discovery loop centered on the LLM as a generative agent, Fig.~\ref{fg:workflow-llm-qc}, inspired by FunSearch~\cite{Fun-Search}.
The framework's primary objective is to discover novel ansatz structures $|\psi\rangle$, rather than merely optimizing the parameters $\theta$ of a fixed ansatz. The VQA cost functional $C_O[\psi(\theta)] = \langle \psi(\theta) | O | \psi (\theta)\rangle$ is therefore used as a critical feedback signal to evaluate each structure's representational capability. Here, $O$ represents a problem-specific operator (e.g., the system's Hamiltonian).  

We initiate the process by providing a set of circuit templates. Either this input can be minimal and generic, prompting the LLM's autonomous discovery, or we can actively guide the LLM by embedding known physical knowledge, such as symmetries~\cite{SM}, directly into the templates.
The discovery loop then iterates (Fig.~\ref{fg:workflow-llm-qc}): the LLM generates novel ansatz candidates by evolving the templates provided. Each candidate is then evaluated; its parameters $\theta$ are optimized via VQA to find its minimal cost $C_O$, which is balanced against circuit complexity to produce a single quality score~\cite{SM}. Based on this feedback, the LLM analyzes and guides the evolution of the entire candidate pool. It applies LLM-driven ``semantic" mutations and crossovers within a genetic algorithm structure~\cite{song2025iteratedagentsymbolicregression,ideasearch2025,Fun-Search,SM} to create the next generation of circuits, efficiently navigating the search space and escaping local optima.

In the following sections, we apply this protocol to benchmark $XY$ spin chain models and a $2$+$1$d scalar field theory, demonstrating its rapid convergence to compact, high-fidelity circuits scalable to larger system sizes.

\textbf{\textit{The $XY$ Model.}}  
We first apply our framework to prepare the ground state of the anisotropic $XY$ spin chain:
\bea 
H_{XY} = \sum_{i}^{n-1}
\left(\frac{1+\gamma}{2}\sigma_i^x\sigma_{i+1}^x
+ \frac{1-\gamma}{2} \sigma_i^y \sigma_{i+1}^y\right)
+ g_z \sum_i^n \sigma_i^z \,, \nonumber \\ 
\eea 
with $n=9$, $g_z = 1.0$,  $\gamma = 1.0$, and open boundary condition. The objective is to minimize the energy expectation $C_{H_{XY}}[\psi(\theta_i)] = \langle \psi(\theta_i) | H_{XY} | \psi (\theta_i)\rangle \equiv E $. 

\begin{figure}[htbp]
   \begin{center}
    \includegraphics[scale=0.15]{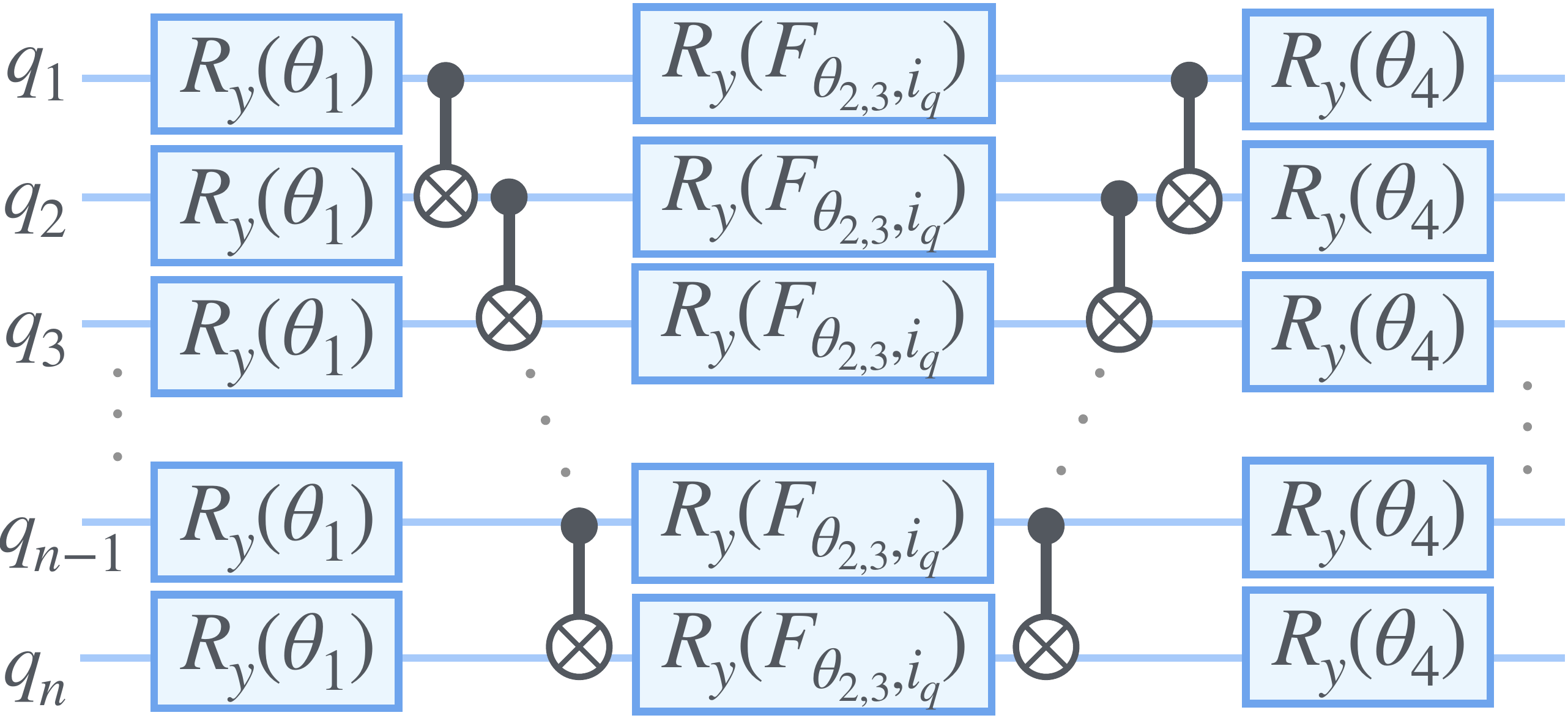} 
 \caption{Best ansatz returned for the $XY$ model.}
   \label{fg:best-llm}
  \end{center}
   \vspace{-5.ex}
 \end{figure}

We initiate the process by providing the framework with only generic circuit templates~\cite{SM}, withholding any expert-crafted ansätze tailored for this problem. The LLM's automated search converged on a highly compact 5-layer, 4-parameter ansatz in Fig.~\ref{fg:best-llm}.

Most notably, the discovered circuit in Fig.~\ref{fg:best-llm}, layer 3, featured a spatially modulated local rotation angle described by
\bea\label{eq:FXY}
F^{\text{AI}}_{\theta_{2,3},i_q} = \theta_2 + \theta_3 \cdot \sin \left( \frac{i}{n}\pi \right) \,,
\eea
where $i$ is the qubit location. 
While this representation may not be perfect, the emergence of this $\sin$ pattern is significant as it breaks translational symmetry, a key feature imposed by the open boundaries, which was not explicitly communicated to the LLM.

Inspired by this LLM-generated physical insight, we collaboratively refine this concept into a more physically grounded function. 
Recognizing that the open boundaries likely cause stronger localization at the edges, we replaced the sinusoidal form with a function that sharpens the profile near the boundaries
\bea
F^{\text{human}}_{\theta_{2,3},i_q} = \theta_2 + \theta_3 \cdot \cos^{2n} \left( \frac{i}{n}\pi \right) \,.
\eea
This form strongly suppresses the rotation in the bulk (where $\cos^{2n} \approx 0$ for large $n$) while maximizing it at the boundaries, better reflecting the physics of a system dominated by boundary effects. 

\begin{figure}
    \centering
    \includegraphics[width=1.02\linewidth]{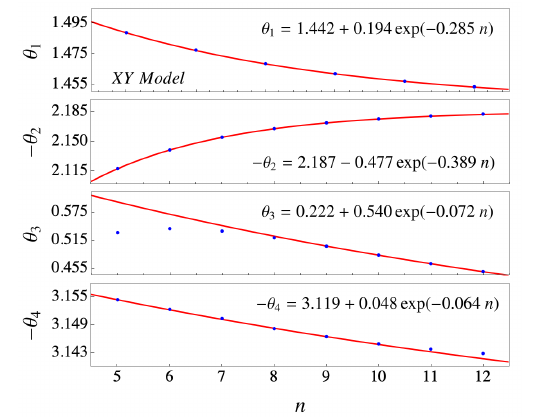}
    \caption{Optimal parameters obtained for the ansatz in Fig. \ref{fg:best-llm} as a function of the system size $n$. 
    The red lines correspond to the fitting functions shown in the legend. }
    \label{fg:theta_XY}
\end{figure}

\begin{figure}
    \centering
    \includegraphics[width=\linewidth]{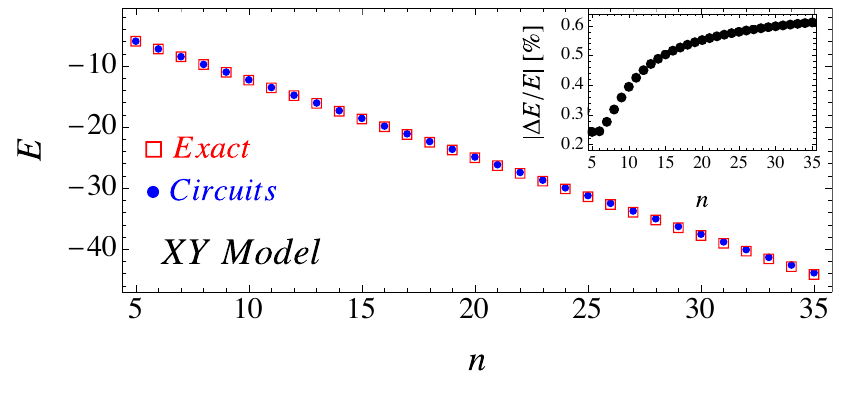}
    \caption{Ground-state energy for different system sizes. Blue dots show energies obtained from circuits using the fitted parameters in Fig.~\ref{fg:theta_XY}, and red squares indicate the exact ground-state energies. The inset shows the relative energy deviation from the exact values.
    }
    \label{fg:energy-XY}
\end{figure}

This final, human-AI-collaborative ansatz achieves remarkable results: a ground state fidelity $F = 0.980$ and an energy error $|\Delta E/E| < 0.4\%$, while remaining compact (4 parameters). This contrasts sharply with typical VQAs (e.g., Ref.~\cite{Xu:2025gwo} used ${\cal O}(20)$ parameters). The compactness of this ansatz enabled robust scalability. We fixed the 4-parameter structure and fit the optimized parameters for small $n$, from $n=4$ to $n=10$, to smooth functions as shown in Fig.~\ref{fg:theta_XY}. These functions were then used to extrapolate the circuit up to $n=35$. As shown in Fig.~\ref{fg:energy-XY}, the extrapolated energies, computed via \texttt{quimb}~\cite{Gray2018}, match the exact ground states, with $|\Delta E/E| < 0.6\%$ across all sizes. 
The single-qubit fidelity \(F_s = \sqrt[n]{F}\) remains \(\gtrsim 0.99\), indicating the high accuracy and fidelity during the extrapolation. 
This $1$+$1$d case demonstrates the power of the ``human-AI" synergy: the LLM catalyzes human intuition by discovering non-trivial physical concepts, and human expertise refines these concepts into high-performance, scalable solutions. This dynamic forms the critical frontier we explore next.

 \textbf{\textit{$2+1$ dimensional quantum field theory.}} 
We now apply our ``human-AI" paradigm to a primary goal for quantum devices: $2$+$1$d QFT simulation. We focus on a self-interacting scalar field theory $\phi(\bold{x})$. We discretize the theory onto an $n \times n$ spatial lattice with periodic boundary conditions using the $n_q = 1$ (minimal nontrivial truncations of the local scalar field degrees of freedom) Jordan-Lee-Preskill framework~\cite{Jordan:2012xnu}.
In the $n_q = 1$ case, the non-trivial interaction term takes the form of $\phi(\bold{x})^3$.
The discretized Hamiltonian is given by
\begin{eqnarray}
\hat{H} &=& \frac{a^2}{2}
\sum_{\bold{x}}
\left[
\hat{\Pi}(\bold{x})^2
- \hat{\phi}(\bold{x})\nabla^2 \hat{\phi}(\bold{x})
+ \frac{\lambda}{3} \hat{\phi}(\bold{x})^3
\right]\notag\\
&\equiv& \hat{H}_k + \hat{H}_\phi +\hat{H}_{\rm int}.
\label{eq:Hamiltonian}
\end{eqnarray}
Further details and definitions of each operator in  Eq.~\ref{eq:Hamiltonian} are provided in the SM~\cite{SM}. 
 
In the following study, we deliberately tuned the truncation parameter to generate a small ground-to-first-excited gap, thereby creating a VQA-hostile environment with large ground-state entanglement entropy.

We note that the digitalized system possesses both translational and ${\rm C}_4$ ($90^\circ$) rotational symmetry.  
Directly inspired by the effectiveness of symmetry observed in the $XY$ model, we instruct the LLM to search exclusively within the symmetry-preserving subspace, instead of relying on the LLM to autonomously discover the symmetry, as it did with the boundary conditions in the $XY$ model.

\begin{figure}
     \centering
     \includegraphics[width=\linewidth]{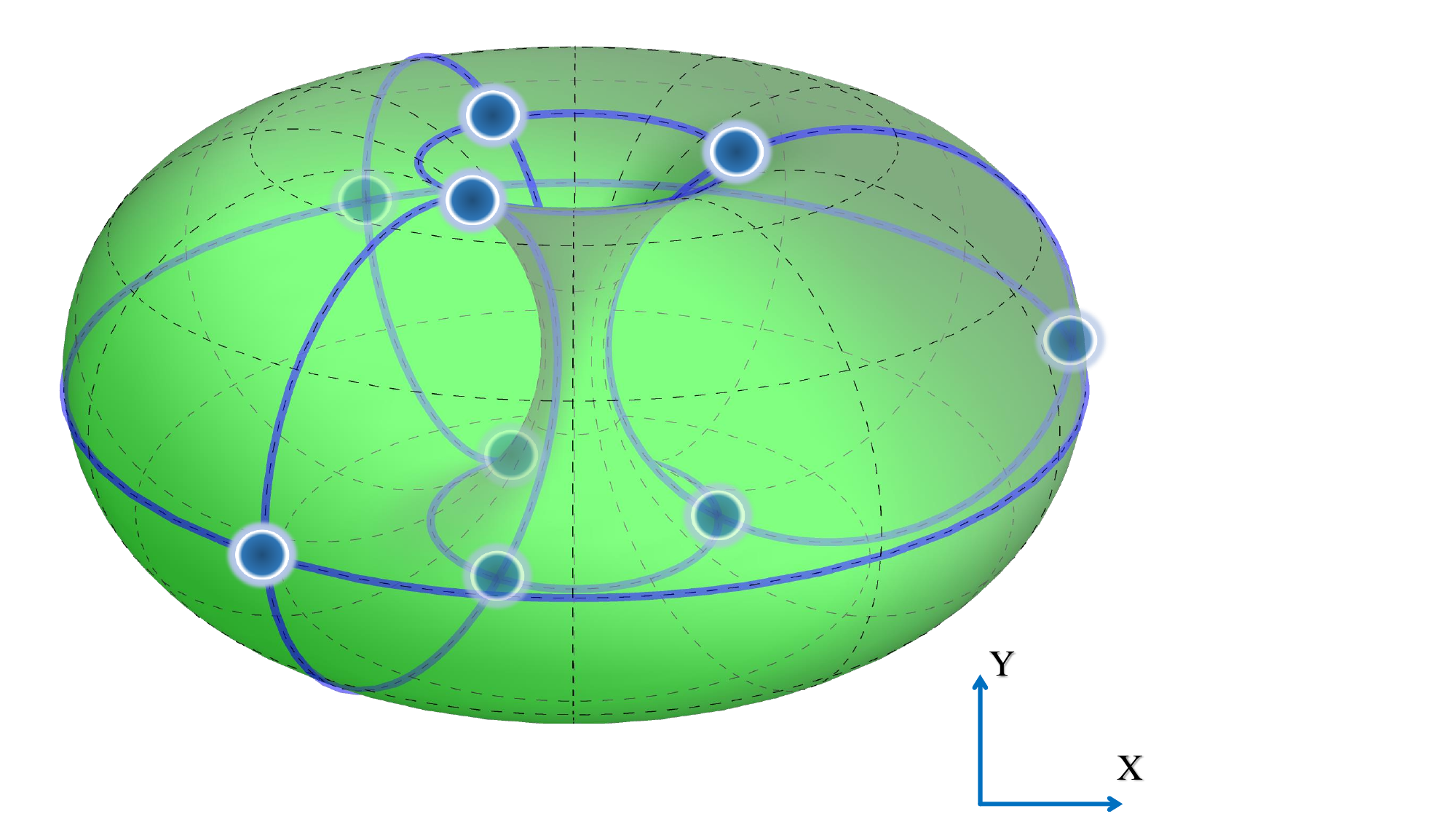}
     \caption{The geometry of the two-qubit meta-circuit $\rm MCZ (\theta)$ for the $2$+$1$d scalar field theory. 
     }
     \label{fig:meta}
 \end{figure}

We implemented this guidance by providing the LLM
with ``meta-circuits” as its building blocks. These blocks
are symmetric by construction. Our two primary meta-circuits are \({\rm MC}_U(\theta)\) and \({\rm MCZ}(\theta)\). The \({\rm MC}_U(\theta)\) applies possible single-qubit gate \(U(\theta)\) to every site, manifestly preserving the symmetry. While, the meta-circuit \({\rm MCZ}(\theta)\) applies the same \({\rm CRZ}(\theta)\) gates sequentially to every adjacent pair. The \({\rm C}_4\) rotational symmetry makes \({\rm CRZ}\) the uniquely viable option for the 2-qubit entangling gate. 
Figure~\ref{fig:meta} illustrates its structure under periodic boundary conditions on the spatial lattice.

\begin{figure}
     \centering
     \includegraphics[width=\linewidth]{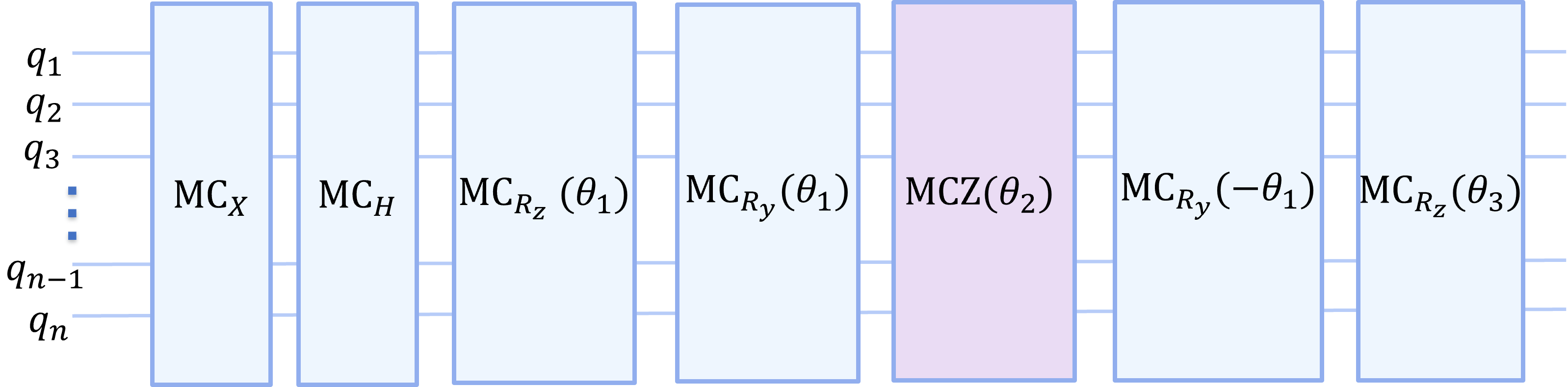}
     \caption{The LLM-discovered ansatz for preparing the ground state of a $2$+$1$d scalar field theory. 
     }
     \label{fg:scalar-circuit}
 \end{figure}

We then employ the LLM to discover the optimal sequence of these physical blocks~\cite{SM}. For a $3\times 3$ system, the framework converged on a remarkably compact ansatz requiring only 3 parameters and 7 meta-circuits (total depth 24 when decomposed with the universal quantum gate), as shown in Fig.~\ref{fg:scalar-circuit}.

This compact ansatz is not just efficient; it is physically scalable. As shown in Table~\ref{tab:parameters}, the VQE energies achieve $<1\%$ error for $n \in \{2, 3, 4, 5\}$. Crucially, the optimal parameters $\theta$ converge to constants for $n \ge 4$. This allowed us to extrapolate the $n=5$ parameters to larger $n=6, 7$ systems. The energies from the extrapolated circuits are consistent with the exact results obtained using adiabatic evolution (Table~\ref{tab:parameters}), verifying the ansatz's scalability.
To our best knowledge, it is the first demonstration of a scalable VQA ansatz for this $2$+$1$d QFT. 

Finally, we benchmarked the $3$-parameter LLM ansatz against a conventional $6$-parameter ``tentative ansatz" 
\bea \label{eq:tentative}
U_{\text{tentative}}(\theta) =
\prod_{\ell=1}^{r}
\left[
e^{-i \hat{H}_\phi \theta_1^{(\ell)}}
e^{-i \hat{H}_{\text{int}} \theta_2^{(\ell)}}
e^{-i \hat{H}_k \theta_3^{(\ell)}}
\right]\,,
\eea 
 and a minimal repetition $r = 2$ is found to be required to achieve a fidelity above $90\%$. 
As shown in Table~\ref{tab:compare-ansatz}, for the comparable high fidelity, the LLM ansatz provides a dramatic reduction in both circuit depth ($24$ vs $72$) and parameter count ($3$ vs $6$). Moreover, the tentative ansatz's parameters showed no clear extrapolative pattern, highlighting the superior, physically-grounded nature of the solution discovered through our guided ``human-AI" collaboration.
\begin{table}[ht]
\centering
\begin{tabular}{lcccccc}
\toprule
$n$ & $E_{\text{VQE}}$ & $E_{0}$ & $F_s$&$\theta_1$ & $\theta_2$ & $\theta_3$ \\
\midrule
2 & 3.84  & 3.78  &0.99& $2.39$ & $0.33$ & $4.32$ \\
3 & 8.79  & 8.74  &0.99& $2.37$         & $0.39$ & $4.40$         \\
4 & 15.69 & 15.60 &0.99& $2.38$ & $0.37$ & $4.37$         \\
5 & 24.52 & 24.39 &0.99& $2.38$ & $0.37$ & $4.38$ \\
6 & $35.32^{+0.02}_{-0.02}$ & $35.33^{+0.02}_{-0.02}$ &-& $2.38$         & $0.37$         & $4.38$         \\
7 & $48.10^{+0.03}_{-0.03}$ & $48.08^{+0.03}_{-0.03}$ &-& $2.38$         & $0.37$         & $4.38$         \\
\bottomrule
\end{tabular}
\caption{Performance of LLM ansatz and optimized parameters for different system $n\times n$. 
$E_{0}$ is the true ground energy, except for $n\ge 6$ where
$E_{0}$ is the ground energy via adiabatic evolution by gradually turning on \(\hat{H}_k\). The adiabatic evolution is performed with a total evolution time \(T = 30\) and time step \(\delta t = 0.05\), a protocol verified to reproduce the \(n = 4\) ground state at 99.8\% precision.
}
\label{tab:parameters}
\end{table}

\begin{table}[ht]
\centering
\begin{tabular}{lcc}
\toprule
Metric & LLM Ansatz & Tentative Ansatz \\
\midrule
Circuit Depth & 24 & 72 \\
Number of Parameters & 3 & 6 \\
$E_{\text{VQE}}$ & 8.794 & 8.771 \\
Fidelity & 0.941 & 0.947 \\
\bottomrule
\end{tabular}
\caption{Comparison of the LLM ansatz and the tentative ansatz in Eq. \ref{eq:tentative} for a $3\times3$ system. }
\label{tab:compare-ansatz}
\end{table}

\begin{figure}
    \centering
    \includegraphics[width=\linewidth]{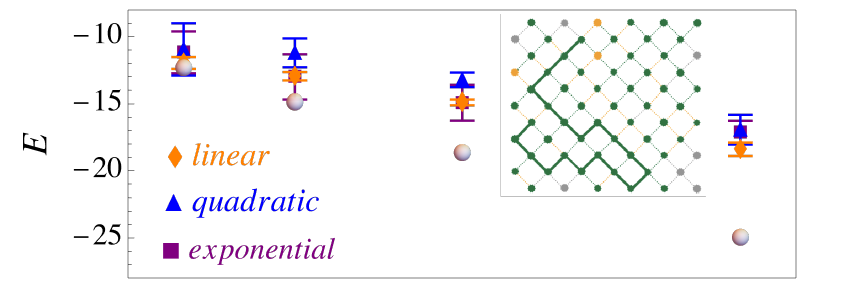}
    \includegraphics[width=\linewidth]{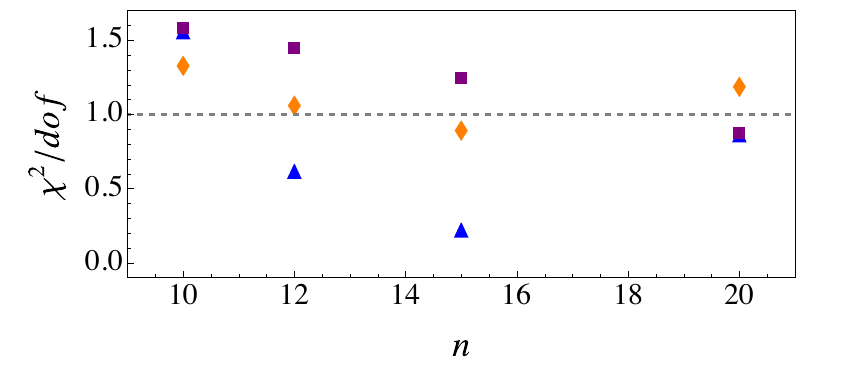}
  \caption{(Top) Extrapolated energy of the XY model states prepared on the Zuchongzhi quantum chip, using exponential (purple square), quadratic (blue triangle), and linear (orange diamond) fits. The white circle shows the result from a noiseless simulator. The chip architecture is shown in the inset. (Bottom) The corresponding $\chi^2$/dof for each fit.}
    \label{fig:zuchongzhi}
\end{figure}

\textbf{\textit{NISQ Hardware Validation.}}
Finally, we validate the practical utility of our compact, LLM-discovered ansätze, which are ideal for the NISQ era due to their minimal parameter count and shallow depth. We executed our $4$-parameter $XY$ model ansatz on the \texttt{Zuchongzhi} quantum chip~\cite{Zhu:2021gkn}, selecting a qubit chain satisfying the singlet-qubit gate fidelity higher than $99.5\%$ and two-qubit gates higher than $98\%$, as shown in Fig.~\ref{fig:zuchongzhi}, top inset. We employed standard Zero-Noise Extrapolation (ZNE)~\cite{Temme:2016vkz,Li:2016vmf} by amplifying the CNOT gate count to mitigate errors. All details regarding ZNE implementation and the comparative analysis of linear, quadratic, and exponential extrapolations are provided in the SM~\cite{SM}.

As shown in Fig.~\ref{fig:zuchongzhi}, the linearly extrapolated energies (our most reliable fit, whereas the covariant matrix for the exponential fit is substantially larger than 1) for $n=10$ agree with the noiseless simulator values within $1\sigma$ uncertainty. While larger systems ($n=20$) show deviations ($\sim30\%$), these are consistent with the expected error accumulation from unmitigated single-qubit gates. This successful execution on noisy hardware, made possible by the 4-parameter simplicity of the LLM ansatz, confirms the real-world advantage of our collaborative framework. A similar demonstration for the conventional ${\cal O}(20)$-parameter ansatz~\cite{Xu:2025gwo} would be computationally intractable, highlighting the practical necessity of discovering physically-compact circuits.

\textbf{\textit{  Conclusion and Outlook.}} 
We have established and validated a human-LLM collaborative framework for quantum circuit design, yielding ansätze that are physically insightful, scalable, and hardware-efficient. This QFT success was built upon the framework's autonomous-discovery capability, manifested for a $1$+$1$d $XY$ model. One of our primary achievements is a $3$-parameter shallow circuit for a $2$+$1$d scalar field theory, which captures the essential challenges for eventually simulating realistic QFTs, such as QCD, without unnecessary complications~\cite{Jordan:2012xnu}. 
Critically, we find the optimal values of the parameters converge to system-size-independent constants for lattices $n \ge 4$, confirming it has captured the essential local physics and thus solving a key scaling bottleneck for VQAs. To our best knowledge, this is the first VQA ansatz to demonstrate this essential scalable property for a 2+1d QFT system. This work demonstrates a practical pathway for AI-assisted, human-guided discovery in scalable quantum simulations. Future research will extend this framework to more geometrically complex systems, including high-dimensional gauge field theories, whose efficient state preparation is a critical component for tackling longstanding challenges in high-energy physics via quantum computing~\cite{Bauer:2022hpo, DiMeglio:2023nsa, np-review, Fang:2024ple,Bauer:2023qgm}.

\begin{acknowledgments}

\textbf{\textit{Acknowledgements.}} 
We thank Xu Feng, Futian Liang, Xiaoyang Wang, Zhi-Cheng Yang, and Xiao Yuan for fruitful discussions. The authors gratefully acknowledge the valuable discussions and insights provided by the members of the Collaboration on Precision Tests and New Physics (CPTNP). The work is partly supported by the National Science Foundation of China under Grant Nos. 12522509, 12235001, 12175016, 12425505, 12305107. This work was also partly supported by Tianyan Quantum Computing Cloud Platform.

\end{acknowledgments}

\bibliography{sample}

\begin{thebibliography}{27}%
\makeatletter
\providecommand \@ifxundefined [1]{%
 \@ifx{#1\undefined}
}%
\providecommand \@ifnum [1]{%
 \ifnum #1\expandafter \@firstoftwo
 \else \expandafter \@secondoftwo
 \fi
}%
\providecommand \@ifx [1]{%
 \ifx #1\expandafter \@firstoftwo
 \else \expandafter \@secondoftwo
 \fi
}%
\providecommand \natexlab [1]{#1}%
\providecommand \enquote  [1]{``#1''}%
\providecommand \bibnamefont  [1]{#1}%
\providecommand \bibfnamefont [1]{#1}%
\providecommand \citenamefont [1]{#1}%
\providecommand \href@noop [0]{\@secondoftwo}%
\providecommand \href [0]{\begingroup \@sanitize@url \@href}%
\providecommand \@href[1]{\@@startlink{#1}\@@href}%
\providecommand \@@href[1]{\endgroup#1\@@endlink}%
\providecommand \@sanitize@url [0]{\catcode `\\12\catcode `\$12\catcode
  `\&12\catcode `\#12\catcode `\^12\catcode `\_12\catcode `\%12\relax}%
\providecommand \@@startlink[1]{}%
\providecommand \@@endlink[0]{}%
\providecommand \url  [0]{\begingroup\@sanitize@url \@url }%
\providecommand \@url [1]{\endgroup\@href {#1}{\urlprefix }}%
\providecommand \urlprefix  [0]{URL }%
\providecommand \Eprint [0]{\href }%
\providecommand \doibase [0]{https://doi.org/}%
\providecommand \selectlanguage [0]{\@gobble}%
\providecommand \bibinfo  [0]{\@secondoftwo}%
\providecommand \bibfield  [0]{\@secondoftwo}%
\providecommand \translation [1]{[#1]}%
\providecommand \BibitemOpen [0]{}%
\providecommand \bibitemStop [0]{}%
\providecommand \bibitemNoStop [0]{.\EOS\space}%
\providecommand \EOS [0]{\spacefactor3000\relax}%
\providecommand \BibitemShut  [1]{\csname bibitem#1\endcsname}%
\let\auto@bib@innerbib\@empty
\bibitem [{\citenamefont {Harrigan}\ \emph {et~al.}(2021)\citenamefont
  {Harrigan} \emph {et~al.}}]{Harrigan:2020dvo}%
  \BibitemOpen
  \bibfield  {author} {\bibinfo {author} {\bibfnamefont {M.~P.}\ \bibnamefont
  {Harrigan}} \emph {et~al.},\ }\bibfield  {title} {\bibinfo {title} {{Quantum
  approximate optimization of non-planar graph problems on a planar
  superconducting processor}},\ }\href
  {https://doi.org/10.1038/s41567-020-01105-y} {\bibfield  {journal} {\bibinfo
  {journal} {Nature Phys.}\ }\textbf {\bibinfo {volume} {17}},\ \bibinfo
  {pages} {332} (\bibinfo {year} {2021})},\ \Eprint
  {https://arxiv.org/abs/2004.04197} {arXiv:2004.04197 [quant-ph]} \BibitemShut
  {NoStop}%
\bibitem [{\citenamefont {Ebadi}\ \emph {et~al.}(2022)\citenamefont {Ebadi}
  \emph {et~al.}}]{Ebadi:2022oxd}%
  \BibitemOpen
  \bibfield  {author} {\bibinfo {author} {\bibfnamefont {S.}~\bibnamefont
  {Ebadi}} \emph {et~al.},\ }\bibfield  {title} {\bibinfo {title} {{Quantum
  Optimization of Maximum Independent Set using Rydberg Atom Arrays}},\ }\href
  {https://doi.org/10.1126/science.abo6587} {\bibfield  {journal} {\bibinfo
  {journal} {Science}\ }\textbf {\bibinfo {volume} {376}},\ \bibinfo {pages}
  {1209} (\bibinfo {year} {2022})},\ \Eprint {https://arxiv.org/abs/2202.09372}
  {arXiv:2202.09372 [quant-ph]} \BibitemShut {NoStop}%
\bibitem [{\citenamefont {Zhu}\ \emph {et~al.}(2025)\citenamefont {Zhu},
  \citenamefont {Zhuang}, \citenamefont {Qian}, \citenamefont {Ma},
  \citenamefont {Liu}, \citenamefont {Ruan},\ and\ \citenamefont
  {Zhou}}]{Zhu:2024yhe}%
  \BibitemOpen
  \bibfield  {author} {\bibinfo {author} {\bibfnamefont {Y.}~\bibnamefont
  {Zhu}}, \bibinfo {author} {\bibfnamefont {W.}~\bibnamefont {Zhuang}},
  \bibinfo {author} {\bibfnamefont {C.}~\bibnamefont {Qian}}, \bibinfo {author}
  {\bibfnamefont {Y.}~\bibnamefont {Ma}}, \bibinfo {author} {\bibfnamefont
  {D.~E.}\ \bibnamefont {Liu}}, \bibinfo {author} {\bibfnamefont
  {M.}~\bibnamefont {Ruan}},\ and\ \bibinfo {author} {\bibfnamefont
  {C.}~\bibnamefont {Zhou}},\ }\bibfield  {title} {\bibinfo {title} {{A novel
  quantum realization of jet clustering in high-energy physics experiments}},\
  }\href {https://doi.org/10.1016/j.scib.2024.12.020} {\bibfield  {journal}
  {\bibinfo  {journal} {Sci. Bull.}\ }\textbf {\bibinfo {volume} {70}},\
  \bibinfo {pages} {460} (\bibinfo {year} {2025})}\BibitemShut {NoStop}%
\bibitem [{\citenamefont {Wecker}\ \emph {et~al.}(2014)\citenamefont {Wecker},
  \citenamefont {Bauer}, \citenamefont {Clark}, \citenamefont {Hastings},\ and\
  \citenamefont {Troyer}}]{PhysRevA.90.022305}%
  \BibitemOpen
  \bibfield  {author} {\bibinfo {author} {\bibfnamefont {D.}~\bibnamefont
  {Wecker}}, \bibinfo {author} {\bibfnamefont {B.}~\bibnamefont {Bauer}},
  \bibinfo {author} {\bibfnamefont {B.~K.}\ \bibnamefont {Clark}}, \bibinfo
  {author} {\bibfnamefont {M.~B.}\ \bibnamefont {Hastings}},\ and\ \bibinfo
  {author} {\bibfnamefont {M.}~\bibnamefont {Troyer}},\ }\bibfield  {title}
  {\bibinfo {title} {Gate-count estimates for performing quantum chemistry on
  small quantum computers},\ }\href
  {https://doi.org/10.1103/PhysRevA.90.022305} {\bibfield  {journal} {\bibinfo
  {journal} {Phys. Rev. A}\ }\textbf {\bibinfo {volume} {90}},\ \bibinfo
  {pages} {022305} (\bibinfo {year} {2014})}\BibitemShut {NoStop}%
\bibitem [{\citenamefont {Feynman}(1982)}]{Feynman:1981tf}%
  \BibitemOpen
  \bibfield  {author} {\bibinfo {author} {\bibfnamefont {R.~P.}\ \bibnamefont
  {Feynman}},\ }\bibfield  {title} {\bibinfo {title} {{Simulating physics with
  computers}},\ }\href {https://doi.org/10.1007/BF02650179} {\bibfield
  {journal} {\bibinfo  {journal} {Int. J. Theor. Phys.}\ }\textbf {\bibinfo
  {volume} {21}},\ \bibinfo {pages} {467} (\bibinfo {year} {1982})}\BibitemShut
  {NoStop}%
\bibitem [{\citenamefont {Jordan}\ \emph {et~al.}(2012)\citenamefont {Jordan},
  \citenamefont {Lee},\ and\ \citenamefont {Preskill}}]{Jordan:2012xnu}%
  \BibitemOpen
  \bibfield  {author} {\bibinfo {author} {\bibfnamefont {S.~P.}\ \bibnamefont
  {Jordan}}, \bibinfo {author} {\bibfnamefont {K.~S.~M.}\ \bibnamefont {Lee}},\
  and\ \bibinfo {author} {\bibfnamefont {J.}~\bibnamefont {Preskill}},\
  }\bibfield  {title} {\bibinfo {title} {{Quantum Algorithms for Quantum Field
  Theories}},\ }\href {https://doi.org/10.1126/science.1217069} {\bibfield
  {journal} {\bibinfo  {journal} {Science}\ }\textbf {\bibinfo {volume}
  {336}},\ \bibinfo {pages} {1130} (\bibinfo {year} {2012})},\ \Eprint
  {https://arxiv.org/abs/1111.3633} {arXiv:1111.3633 [quant-ph]} \BibitemShut
  {NoStop}%
\bibitem [{\citenamefont {Bauer}\ \emph
  {et~al.}(2023{\natexlab{a}})\citenamefont {Bauer} \emph
  {et~al.}}]{Bauer:2022hpo}%
  \BibitemOpen
  \bibfield  {author} {\bibinfo {author} {\bibfnamefont {C.~W.}\ \bibnamefont
  {Bauer}} \emph {et~al.},\ }\bibfield  {title} {\bibinfo {title} {{Quantum
  Simulation for High-Energy Physics}},\ }\href
  {https://doi.org/10.1103/PRXQuantum.4.027001} {\bibfield  {journal} {\bibinfo
   {journal} {PRX Quantum}\ }\textbf {\bibinfo {volume} {4}},\ \bibinfo {pages}
  {027001} (\bibinfo {year} {2023}{\natexlab{a}})},\ \Eprint
  {https://arxiv.org/abs/2204.03381} {arXiv:2204.03381 [quant-ph]} \BibitemShut
  {NoStop}%
\bibitem [{\citenamefont {Di~Meglio}\ \emph {et~al.}(2024)\citenamefont
  {Di~Meglio} \emph {et~al.}}]{DiMeglio:2023nsa}%
  \BibitemOpen
  \bibfield  {author} {\bibinfo {author} {\bibfnamefont {A.}~\bibnamefont
  {Di~Meglio}} \emph {et~al.},\ }\bibfield  {title} {\bibinfo {title} {{Quantum
  Computing for High-Energy Physics: State of the Art and Challenges}},\ }\href
  {https://doi.org/10.1103/PRXQuantum.5.037001} {\bibfield  {journal} {\bibinfo
   {journal} {PRX Quantum}\ }\textbf {\bibinfo {volume} {5}},\ \bibinfo {pages}
  {037001} (\bibinfo {year} {2024})},\ \Eprint
  {https://arxiv.org/abs/2307.03236} {arXiv:2307.03236 [quant-ph]} \BibitemShut
  {NoStop}%
\bibitem [{\citenamefont {Bauer}\ \emph
  {et~al.}(2023{\natexlab{b}})\citenamefont {Bauer}, \citenamefont {Davoudi},
  \citenamefont {Klco},\ and\ \citenamefont {Savage}}]{np-review}%
  \BibitemOpen
  \bibfield  {author} {\bibinfo {author} {\bibfnamefont {C.~W.}\ \bibnamefont
  {Bauer}}, \bibinfo {author} {\bibfnamefont {Z.}~\bibnamefont {Davoudi}},
  \bibinfo {author} {\bibfnamefont {N.}~\bibnamefont {Klco}},\ and\ \bibinfo
  {author} {\bibfnamefont {M.~J.}\ \bibnamefont {Savage}},\ }\bibfield  {title}
  {\bibinfo {title} {Quantum simulation of fundamental particles and forces},\
  }\href {https://doi.org/10.1038/s42254-023-00599-8} {\bibfield  {journal}
  {\bibinfo  {journal} {Nature Reviews Physics}\ }\textbf {\bibinfo {volume}
  {5}},\ \bibinfo {pages} {420} (\bibinfo {year}
  {2023}{\natexlab{b}})}\BibitemShut {NoStop}%
\bibitem [{\citenamefont {Fang}\ \emph {et~al.}(2025)\citenamefont {Fang},
  \citenamefont {Gao}, \citenamefont {Li}, \citenamefont {Shu}, \citenamefont
  {Wu}, \citenamefont {Xing}, \citenamefont {Xu}, \citenamefont {Xu},\ and\
  \citenamefont {Zhou}}]{Fang:2024ple}%
  \BibitemOpen
  \bibfield  {author} {\bibinfo {author} {\bibfnamefont {Y.}~\bibnamefont
  {Fang}}, \bibinfo {author} {\bibfnamefont {C.}~\bibnamefont {Gao}}, \bibinfo
  {author} {\bibfnamefont {Y.-Y.}\ \bibnamefont {Li}}, \bibinfo {author}
  {\bibfnamefont {J.}~\bibnamefont {Shu}}, \bibinfo {author} {\bibfnamefont
  {Y.}~\bibnamefont {Wu}}, \bibinfo {author} {\bibfnamefont {H.}~\bibnamefont
  {Xing}}, \bibinfo {author} {\bibfnamefont {B.}~\bibnamefont {Xu}}, \bibinfo
  {author} {\bibfnamefont {L.}~\bibnamefont {Xu}},\ and\ \bibinfo {author}
  {\bibfnamefont {C.}~\bibnamefont {Zhou}},\ }\bibfield  {title} {\bibinfo
  {title} {{Quantum frontiers in high energy physics}},\ }\href
  {https://doi.org/10.1007/s11433-024-2635-4} {\bibfield  {journal} {\bibinfo
  {journal} {Sci. China Phys. Mech. Astron.}\ }\textbf {\bibinfo {volume}
  {68}},\ \bibinfo {pages} {260301} (\bibinfo {year} {2025})},\ \Eprint
  {https://arxiv.org/abs/2411.11294} {arXiv:2411.11294 [hep-ph]} \BibitemShut
  {NoStop}%
\bibitem [{\citenamefont {Preskill}(2018)}]{Preskill:2018jim}%
  \BibitemOpen
  \bibfield  {author} {\bibinfo {author} {\bibfnamefont {J.}~\bibnamefont
  {Preskill}},\ }\bibfield  {title} {\bibinfo {title} {{Quantum Computing in
  the NISQ era and beyond}},\ }\href {https://doi.org/10.22331/q-2018-08-06-79}
  {\bibfield  {journal} {\bibinfo  {journal} {Quantum}\ }\textbf {\bibinfo
  {volume} {2}},\ \bibinfo {pages} {79} (\bibinfo {year} {2018})},\ \Eprint
  {https://arxiv.org/abs/1801.00862} {arXiv:1801.00862 [quant-ph]} \BibitemShut
  {NoStop}%
\bibitem [{\citenamefont {McClean}\ \emph {et~al.}(2016)\citenamefont
  {McClean}, \citenamefont {Romero}, \citenamefont {Babbush},\ and\
  \citenamefont {Aspuru-Guzik}}]{McClean:2015vup}%
  \BibitemOpen
  \bibfield  {author} {\bibinfo {author} {\bibfnamefont {J.~R.}\ \bibnamefont
  {McClean}}, \bibinfo {author} {\bibfnamefont {J.}~\bibnamefont {Romero}},
  \bibinfo {author} {\bibfnamefont {R.}~\bibnamefont {Babbush}},\ and\ \bibinfo
  {author} {\bibfnamefont {A.}~\bibnamefont {Aspuru-Guzik}},\ }\bibfield
  {title} {\bibinfo {title} {{The theory of variational hybrid
  quantum-classical algorithms}},\ }\href
  {https://doi.org/10.1088/1367-2630/18/2/023023} {\bibfield  {journal}
  {\bibinfo  {journal} {New J. Phys.}\ }\textbf {\bibinfo {volume} {18}},\
  \bibinfo {pages} {023023} (\bibinfo {year} {2016})},\ \Eprint
  {https://arxiv.org/abs/1509.04279} {arXiv:1509.04279 [quant-ph]} \BibitemShut
  {NoStop}%
\bibitem [{\citenamefont {Farhi}\ \emph {et~al.}(2014)\citenamefont {Farhi},
  \citenamefont {Goldstone},\ and\ \citenamefont {Gutmann}}]{Farhi:2014ych}%
  \BibitemOpen
  \bibfield  {author} {\bibinfo {author} {\bibfnamefont {E.}~\bibnamefont
  {Farhi}}, \bibinfo {author} {\bibfnamefont {J.}~\bibnamefont {Goldstone}},\
  and\ \bibinfo {author} {\bibfnamefont {S.}~\bibnamefont {Gutmann}},\
  }\bibfield  {title} {\bibinfo {title} {{A Quantum Approximate Optimization
  Algorithm}},\ }\href@noop {} {\  (\bibinfo {year} {2014})},\ \Eprint
  {https://arxiv.org/abs/1411.4028} {arXiv:1411.4028 [quant-ph]} \BibitemShut
  {NoStop}%
\bibitem [{\citenamefont {Cerezo}\ \emph {et~al.}(2021)\citenamefont {Cerezo}
  \emph {et~al.}}]{Cerezo:2020jpv}%
  \BibitemOpen
  \bibfield  {author} {\bibinfo {author} {\bibfnamefont {M.}~\bibnamefont
  {Cerezo}} \emph {et~al.},\ }\bibfield  {title} {\bibinfo {title}
  {{Variational quantum algorithms}},\ }\href
  {https://doi.org/10.1038/s42254-021-00348-9} {\bibfield  {journal} {\bibinfo
  {journal} {Nature Rev. Phys.}\ }\textbf {\bibinfo {volume} {3}},\ \bibinfo
  {pages} {625} (\bibinfo {year} {2021})},\ \Eprint
  {https://arxiv.org/abs/2012.09265} {arXiv:2012.09265 [quant-ph]} \BibitemShut
  {NoStop}%
\bibitem [{\citenamefont {McClean}\ \emph {et~al.}(2018)\citenamefont
  {McClean}, \citenamefont {Boixo}, \citenamefont {Smelyanskiy}, \citenamefont
  {Babbush},\ and\ \citenamefont {Neven}}]{McClean:2018jps}%
  \BibitemOpen
  \bibfield  {author} {\bibinfo {author} {\bibfnamefont {J.~R.}\ \bibnamefont
  {McClean}}, \bibinfo {author} {\bibfnamefont {S.}~\bibnamefont {Boixo}},
  \bibinfo {author} {\bibfnamefont {V.~N.}\ \bibnamefont {Smelyanskiy}},
  \bibinfo {author} {\bibfnamefont {R.}~\bibnamefont {Babbush}},\ and\ \bibinfo
  {author} {\bibfnamefont {H.}~\bibnamefont {Neven}},\ }\bibfield  {title}
  {\bibinfo {title} {{Barren plateaus in quantum neural network training
  landscapes}},\ }\href {https://doi.org/10.1038/s41467-018-07090-4} {\bibfield
   {journal} {\bibinfo  {journal} {Nature Commun.}\ }\textbf {\bibinfo {volume}
  {9}},\ \bibinfo {pages} {4812} (\bibinfo {year} {2018})},\ \Eprint
  {https://arxiv.org/abs/1803.11173} {arXiv:1803.11173 [quant-ph]} \BibitemShut
  {NoStop}%
\bibitem [{\citenamefont {Martyniuk}\ \emph {et~al.}(2024)\citenamefont
  {Martyniuk}, \citenamefont {Jung},\ and\ \citenamefont
  {Paschke}}]{Martyniuk:2024bzu}%
  \BibitemOpen
  \bibfield  {author} {\bibinfo {author} {\bibfnamefont {D.}~\bibnamefont
  {Martyniuk}}, \bibinfo {author} {\bibfnamefont {J.}~\bibnamefont {Jung}},\
  and\ \bibinfo {author} {\bibfnamefont {A.}~\bibnamefont {Paschke}},\
  }\bibfield  {title} {\bibinfo {title} {{Quantum Architecture Search: A
  Survey}},\ }in\ \href {https://doi.org/10.1109/QCE60285.2024.00198} {\emph
  {\bibinfo {booktitle} {{2024 International Conference on Quantum Computing
  and Engineering}}}}\ (\bibinfo {year} {2024})\ \Eprint
  {https://arxiv.org/abs/2406.06210} {arXiv:2406.06210 [quant-ph]} \BibitemShut
  {NoStop}%
\bibitem [{\citenamefont {Collaboration}(2025)}]{ideasearch2025}%
  \BibitemOpen
  \bibfield  {author} {\bibinfo {author} {\bibfnamefont {I.}~\bibnamefont
  {Collaboration}},\ }\href@noop {} {\bibinfo {title} {Ideasearch}},\ \bibinfo
  {howpublished} {\url{https://github.com/IdeaSearch/IdeaSearch-fit}} (\bibinfo
  {year} {2025})\BibitemShut {NoStop}%
\bibitem [{\citenamefont {Romera-Paredes}\ \emph {et~al.}(2024)\citenamefont
  {Romera-Paredes}, \citenamefont {Barekatain}, \citenamefont {Novikov},
  \citenamefont {Balog}, \citenamefont {Kumar}, \citenamefont {Dupont},
  \citenamefont {Ruiz}, \citenamefont {Ellenberg}, \citenamefont {Wang},
  \citenamefont {Fawzi}, \citenamefont {Kohli},\ and\ \citenamefont
  {Fawzi}}]{Fun-Search}%
  \BibitemOpen
  \bibfield  {author} {\bibinfo {author} {\bibfnamefont {B.}~\bibnamefont
  {Romera-Paredes}}, \bibinfo {author} {\bibfnamefont {M.}~\bibnamefont
  {Barekatain}}, \bibinfo {author} {\bibfnamefont {A.}~\bibnamefont {Novikov}},
  \bibinfo {author} {\bibfnamefont {M.}~\bibnamefont {Balog}}, \bibinfo
  {author} {\bibfnamefont {M.~P.}\ \bibnamefont {Kumar}}, \bibinfo {author}
  {\bibfnamefont {E.}~\bibnamefont {Dupont}}, \bibinfo {author} {\bibfnamefont
  {F.~J.~R.}\ \bibnamefont {Ruiz}}, \bibinfo {author} {\bibfnamefont {J.~S.}\
  \bibnamefont {Ellenberg}}, \bibinfo {author} {\bibfnamefont {P.}~\bibnamefont
  {Wang}}, \bibinfo {author} {\bibfnamefont {O.}~\bibnamefont {Fawzi}},
  \bibinfo {author} {\bibfnamefont {P.}~\bibnamefont {Kohli}},\ and\ \bibinfo
  {author} {\bibfnamefont {A.}~\bibnamefont {Fawzi}},\ }\bibfield  {title}
  {\bibinfo {title} {Mathematical discoveries from program search with large
  language models},\ }\href@noop {} {\bibfield  {journal} {\bibinfo  {journal}
  {Nature}\ }\textbf {\bibinfo {volume} {625}},\ \bibinfo {pages} {468}
  (\bibinfo {year} {2024})}\BibitemShut {NoStop}%
\bibitem [{SM(2025)}]{SM}%
  \BibitemOpen
  \href@noop {} {\bibinfo {title} {Supplemental material for ``quantum state
  preparation via llm-driven evolution''}} (\bibinfo {year} {2025})\BibitemShut
  {NoStop}%
\bibitem [{\citenamefont {et.
  al.}(2025)}]{song2025iteratedagentsymbolicregression}%
  \BibitemOpen
  \bibfield  {author} {\bibinfo {author} {\bibfnamefont {Z.-Y.~S.}\
  \bibnamefont {et. al.}},\ }\href {https://arxiv.org/abs/2510.08317} {\bibinfo
  {title} {Iterated agent for symbolic regression}} (\bibinfo {year} {2025}),\
  \Eprint {https://arxiv.org/abs/2510.08317} {arXiv:2510.08317
  [physics.comp-ph]} \BibitemShut {NoStop}%
\bibitem [{\citenamefont {Xu}\ \emph {et~al.}(2025)\citenamefont {Xu},
  \citenamefont {Xiao}, \citenamefont {Huang}, \citenamefont {He},
  \citenamefont {Fan},\ and\ \citenamefont {Zeng}}]{Xu:2025gwo}%
  \BibitemOpen
  \bibfield  {author} {\bibinfo {author} {\bibfnamefont {H.}~\bibnamefont
  {Xu}}, \bibinfo {author} {\bibfnamefont {T.}~\bibnamefont {Xiao}}, \bibinfo
  {author} {\bibfnamefont {J.}~\bibnamefont {Huang}}, \bibinfo {author}
  {\bibfnamefont {M.}~\bibnamefont {He}}, \bibinfo {author} {\bibfnamefont
  {J.}~\bibnamefont {Fan}},\ and\ \bibinfo {author} {\bibfnamefont
  {G.}~\bibnamefont {Zeng}},\ }\bibfield  {title} {\bibinfo {title} {{Toward
  Heisenberg Limit without Critical Slowing Down via Quantum Reinforcement
  Learning}},\ }\href {https://doi.org/10.1103/PhysRevLett.134.120803}
  {\bibfield  {journal} {\bibinfo  {journal} {Phys. Rev. Lett.}\ }\textbf
  {\bibinfo {volume} {134}},\ \bibinfo {pages} {120803} (\bibinfo {year}
  {2025})},\ \Eprint {https://arxiv.org/abs/2503.02210} {arXiv:2503.02210
  [quant-ph]} \BibitemShut {NoStop}%
\bibitem [{\citenamefont {Gray}(2018)}]{Gray2018}%
  \BibitemOpen
  \bibfield  {author} {\bibinfo {author} {\bibfnamefont {J.}~\bibnamefont
  {Gray}},\ }\bibfield  {title} {\bibinfo {title} {quimb: A python package for
  quantum information and many-body calculations},\ }\href
  {https://doi.org/10.21105/joss.00819} {\bibfield  {journal} {\bibinfo
  {journal} {Journal of Open Source Software}\ }\textbf {\bibinfo {volume}
  {3}},\ \bibinfo {pages} {819} (\bibinfo {year} {2018})}\BibitemShut {NoStop}%
\bibitem [{\citenamefont {Zhu}\ \emph {et~al.}(2022)\citenamefont {Zhu} \emph
  {et~al.}}]{Zhu:2021gkn}%
  \BibitemOpen
  \bibfield  {author} {\bibinfo {author} {\bibfnamefont {Q.}~\bibnamefont
  {Zhu}} \emph {et~al.},\ }\bibfield  {title} {\bibinfo {title} {{Quantum
  computational advantage via 60-qubit 24-cycle random circuit sampling}},\
  }\href {https://doi.org/10.1016/j.scib.2021.10.017} {\bibfield  {journal}
  {\bibinfo  {journal} {Sci. Bull.}\ }\textbf {\bibinfo {volume} {67}},\
  \bibinfo {pages} {240} (\bibinfo {year} {2022})},\ \Eprint
  {https://arxiv.org/abs/2109.03494} {arXiv:2109.03494 [quant-ph]} \BibitemShut
  {NoStop}%
\bibitem [{\citenamefont {Temme}\ \emph {et~al.}(2017)\citenamefont {Temme},
  \citenamefont {Bravyi},\ and\ \citenamefont {Gambetta}}]{Temme:2016vkz}%
  \BibitemOpen
  \bibfield  {author} {\bibinfo {author} {\bibfnamefont {K.}~\bibnamefont
  {Temme}}, \bibinfo {author} {\bibfnamefont {S.}~\bibnamefont {Bravyi}},\ and\
  \bibinfo {author} {\bibfnamefont {J.~M.}\ \bibnamefont {Gambetta}},\
  }\bibfield  {title} {\bibinfo {title} {{Error Mitigation for Short-Depth
  Quantum Circuits}},\ }\href {https://doi.org/10.1103/physrevlett.119.180509}
  {\bibfield  {journal} {\bibinfo  {journal} {Phys. Rev. Lett.}\ }\textbf
  {\bibinfo {volume} {119}},\ \bibinfo {pages} {180509} (\bibinfo {year}
  {2017})},\ \Eprint {https://arxiv.org/abs/1612.02058} {arXiv:1612.02058
  [quant-ph]} \BibitemShut {NoStop}%
\bibitem [{\citenamefont {Li}\ and\ \citenamefont
  {Benjamin}(2017)}]{Li:2016vmf}%
  \BibitemOpen
  \bibfield  {author} {\bibinfo {author} {\bibfnamefont {Y.}~\bibnamefont
  {Li}}\ and\ \bibinfo {author} {\bibfnamefont {S.~C.}\ \bibnamefont
  {Benjamin}},\ }\bibfield  {title} {\bibinfo {title} {{Efficient Variational
  Quantum Simulator Incorporating Active Error Minimization}},\ }\href
  {https://doi.org/10.1103/physrevx.7.021050} {\bibfield  {journal} {\bibinfo
  {journal} {Phys. Rev. X}\ }\textbf {\bibinfo {volume} {7}},\ \bibinfo {pages}
  {021050} (\bibinfo {year} {2017})},\ \Eprint
  {https://arxiv.org/abs/1611.09301} {arXiv:1611.09301 [quant-ph]} \BibitemShut
  {NoStop}%
\bibitem [{\citenamefont {Bauer}\ \emph
  {et~al.}(2023{\natexlab{c}})\citenamefont {Bauer}, \citenamefont {Davoudi},
  \citenamefont {Klco},\ and\ \citenamefont {Savage}}]{Bauer:2023qgm}%
  \BibitemOpen
  \bibfield  {author} {\bibinfo {author} {\bibfnamefont {C.~W.}\ \bibnamefont
  {Bauer}}, \bibinfo {author} {\bibfnamefont {Z.}~\bibnamefont {Davoudi}},
  \bibinfo {author} {\bibfnamefont {N.}~\bibnamefont {Klco}},\ and\ \bibinfo
  {author} {\bibfnamefont {M.~J.}\ \bibnamefont {Savage}},\ }\bibfield  {title}
  {\bibinfo {title} {{Quantum simulation of fundamental particles and
  forces}},\ }\href {https://doi.org/10.1038/s42254-023-00599-8} {\bibfield
  {journal} {\bibinfo  {journal} {Nature Rev. Phys.}\ }\textbf {\bibinfo
  {volume} {5}},\ \bibinfo {pages} {420} (\bibinfo {year}
  {2023}{\natexlab{c}})},\ \Eprint {https://arxiv.org/abs/2404.06298}
  {arXiv:2404.06298 [hep-ph]} \BibitemShut {NoStop}%
\bibitem [{\citenamefont {Klco}\ and\ \citenamefont
  {Savage}(2019)}]{Klco_2019}%
  \BibitemOpen
  \bibfield  {author} {\bibinfo {author} {\bibfnamefont {N.}~\bibnamefont
  {Klco}}\ and\ \bibinfo {author} {\bibfnamefont {M.~J.}\ \bibnamefont
  {Savage}},\ }\bibfield  {title} {\bibinfo {title} {Digitization of scalar
  fields for quantum computing},\ }\bibfield  {journal} {\bibinfo  {journal}
  {Physical Review A}\ }\textbf {\bibinfo {volume} {99}},\ \href
  {https://doi.org/10.1103/physreva.99.052335} {10.1103/physreva.99.052335}
  (\bibinfo {year} {2019})\BibitemShut {NoStop}%
\end{thebibliography}%

\appendix

\clearpage

\begin{widetext}
\section{Supplemental Material for ``Quantum State Preparation via LLM-Driven Evolution"}

\subsection{A. Additional Details on the scalar field}
In the Jordan-Lee-Preskill framework ~\cite{Jordan:2012xnu}, the scalar field $\phi$ is first discretized onto an $n\times n$ spatial lattice with lattice spacing $a$, after which the local value $\phi(\mathbf{x})$ at each site is digitized by encoding it into $n_q$ qubits, representing $2^{n_q}$ possible discrete values. For our demonstration, we choose $n_q = 1$ to minimize the number of local degrees of freedom. This results in the field $\phi$ being digitized into two values $\{-\phi_{\rm max}, \phi_{\rm max}\}$. Consequently, the second-quantized field operator $\hat{\phi}(\mathbf{x})$ and its canonical momentum operator $\hat{\Pi}(\mathbf{x})$ at site $\mathbf{x}$ are given by \cite{Klco_2019} 
\begin{equation}
\hat{\phi}(\bold{x}){}
= \phi_{\rm max}
\begin{pmatrix}
1 & 0 \\
0 &  -1
\end{pmatrix}, \,
\hat{\Pi}(\bold{x})^2
= \frac{1}{{4\phi^2_{\rm max}}}
\begin{pmatrix}
2 & 1 \\
1 &  2
\end{pmatrix}.
\end{equation} 
where the $+1$ terms in the off-diagonal corners of the operator $\hat{\Pi}(\bold{x})^2$ is due to the choice of a twisted boundary condition in the local field space.
In the $n_q = 1$ case, $\hat{\phi}(\mathbf{x})^2$ is proportional to the identity, rendering the mass term trivial and will not be included. This also leads us to introduce a non-trivial interaction term of the form $\hat{\phi}(\mathbf{x})^3$ with strength $\lambda$.
The discretized Hamiltonian is thus given by
\begin{eqnarray}
\hat{H} &=& \frac{a^2}{2}
\sum_{\bold{x}}
\left[
\hat{\Pi}(\bold{x})^2
- \hat{\phi}(\bold{x})\nabla^2 \hat{\phi}(\bold{x})
+ \frac{\lambda}{3} \hat{\phi}(\bold{x})^3
\right]\notag\\
&\equiv& \hat{H}_k + \hat{H}_\phi +\hat{H}_{\rm int}.
\label{eq:Hamiltonian}
\end{eqnarray}
$\nabla^2$ is the lattice Laplacian operator acting on the field operator, with periodic boundary conditions imposed along both spatial directions.

We set the benchmark coupling at \(\lambda = 0.2\) and lattice spacing \(a = 1\), expressing all dimensional quantities in units of \(a\). The parameter \(\phi_{\text{max}}\) is selected not to minimize digitization errors in the ground state energy \cite{Klco_2019}, but to make ground state preparation nontrivial by targeting a small first excited-state energy gap and large ground-state entanglement entropy. 
To clarify the rationale for our selection of $\phi_\text{max}$, we present heatmaps of the energy gap $\Delta E$ between the first excited state and the ground state and the bipartition entanglement entropy $S_B$ on the $\lambda-\phi_\text{max}$ plane. 
The entanglement entropy is defined for the true ground state $\ket{\psi_{GS}}$ of the system, which is obtained via diagonalization of $\hat{H}$, given in Eq.~\eqref{eq:Hamiltonian}. To define the partition, the $N=n^2$ qubits on the $n \times n$ lattice are indexed sequentially in a row-major (raster-scan) order. This ordering proceeds from left to right across each row, starting from the top row. We employ a spatial bipartition, dividing the system into subsystem $A$ (comprising the first $\lfloor n^2/2 \rfloor$ qubits based on this index) and subsystem $B$ (the remainder). This division effectively bisects the spatial lattice into two contiguous regions (e.g., a top half and a bottom half). The reduced density matrix for subsystem $B$ is computed by tracing out subsystem $A$, $\rho_B = \text{Tr}_A(|\psi_{GS}\rangle\langle\psi_{GS}|)$. The entanglement entropy is then quantified by the von Neumann entropy:
\begin{equation}
    S_B=-\text{Tr}(\rho_B\ln(\rho_B))
    \label{eq:entanglement-entropy}
\end{equation}

As shown in Fig.~\ref{fig:heatmap}, around $\phi_\text{max} \approx 0.45$ for $\lambda = 0.2$, the system exhibits both a small energy gap $\Delta E$ (top row) and large bipartite entanglement entropy $S_B$ (bottom row). This combination defines a regime that is particularly challenging for ground state preparation in quantum simulations.
\begin{figure}[htbp]
\centering

\begin{subfigure}{0.48\textwidth}
\centering
\includegraphics[width=\linewidth]{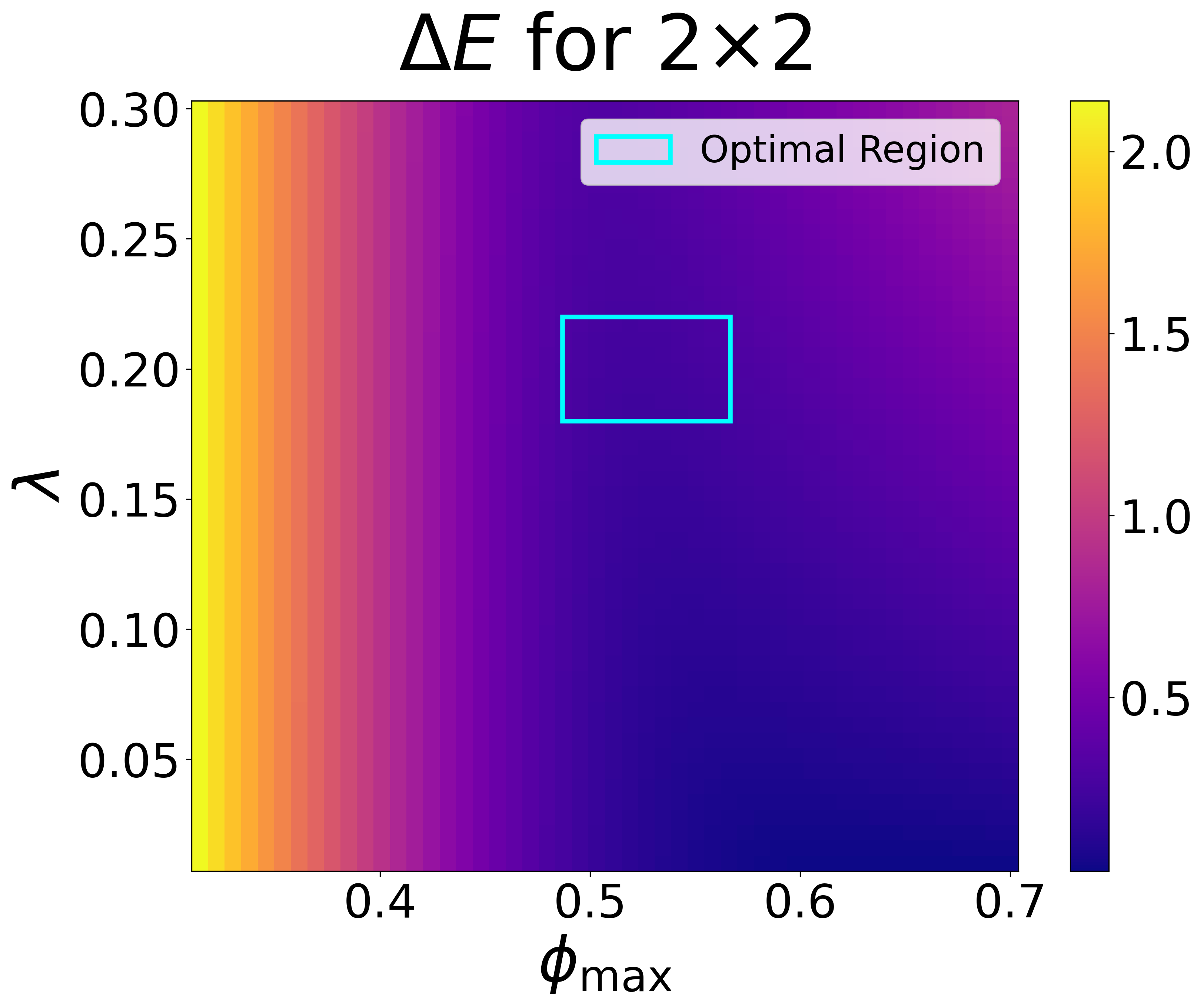}
\end{subfigure}\hfill
\begin{subfigure}{0.48\textwidth}
\centering
\includegraphics[width=\linewidth]{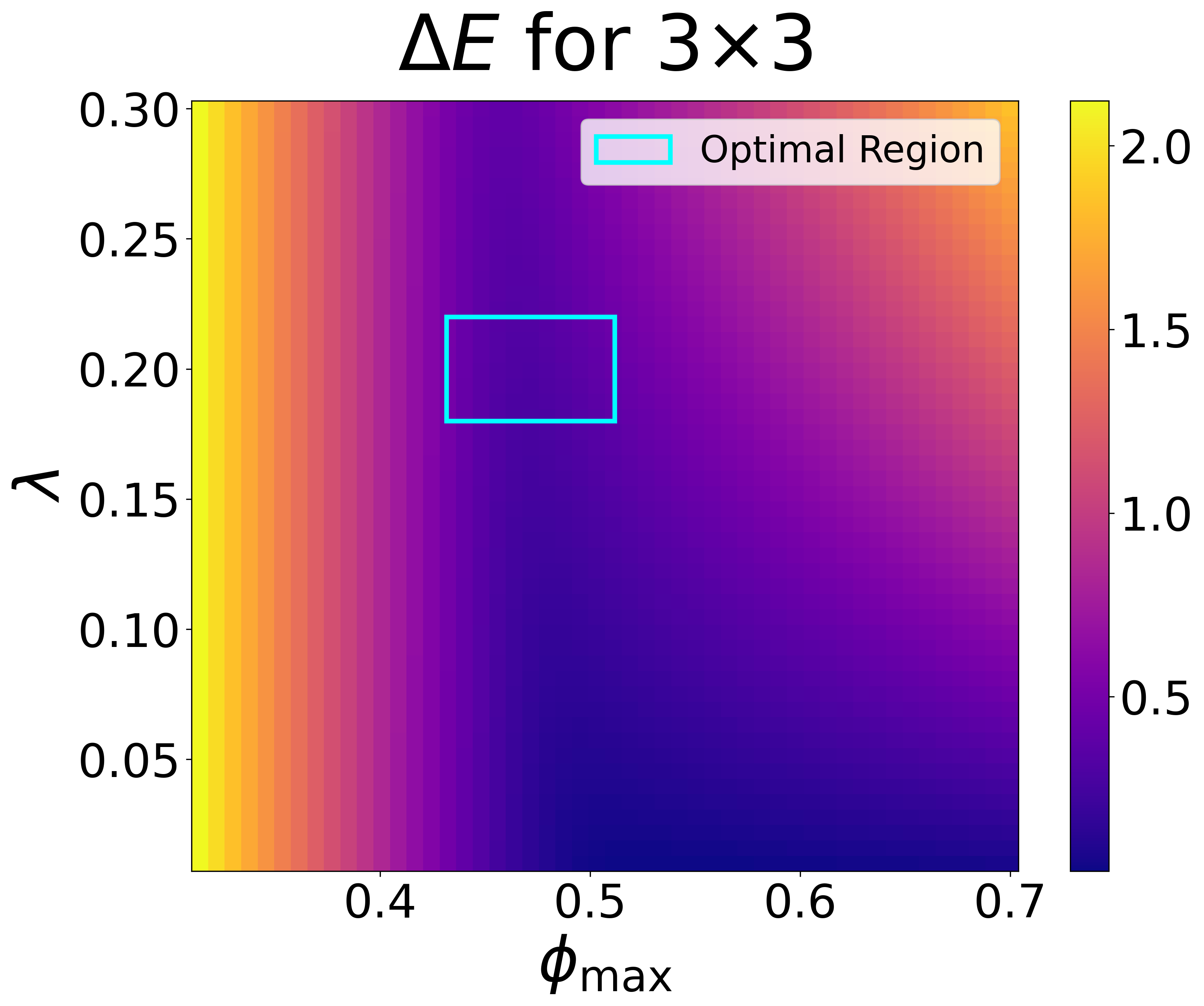}
\end{subfigure}

\vspace{0.6em}

\begin{subfigure}{0.48\textwidth}
\centering
\includegraphics[width=\linewidth]{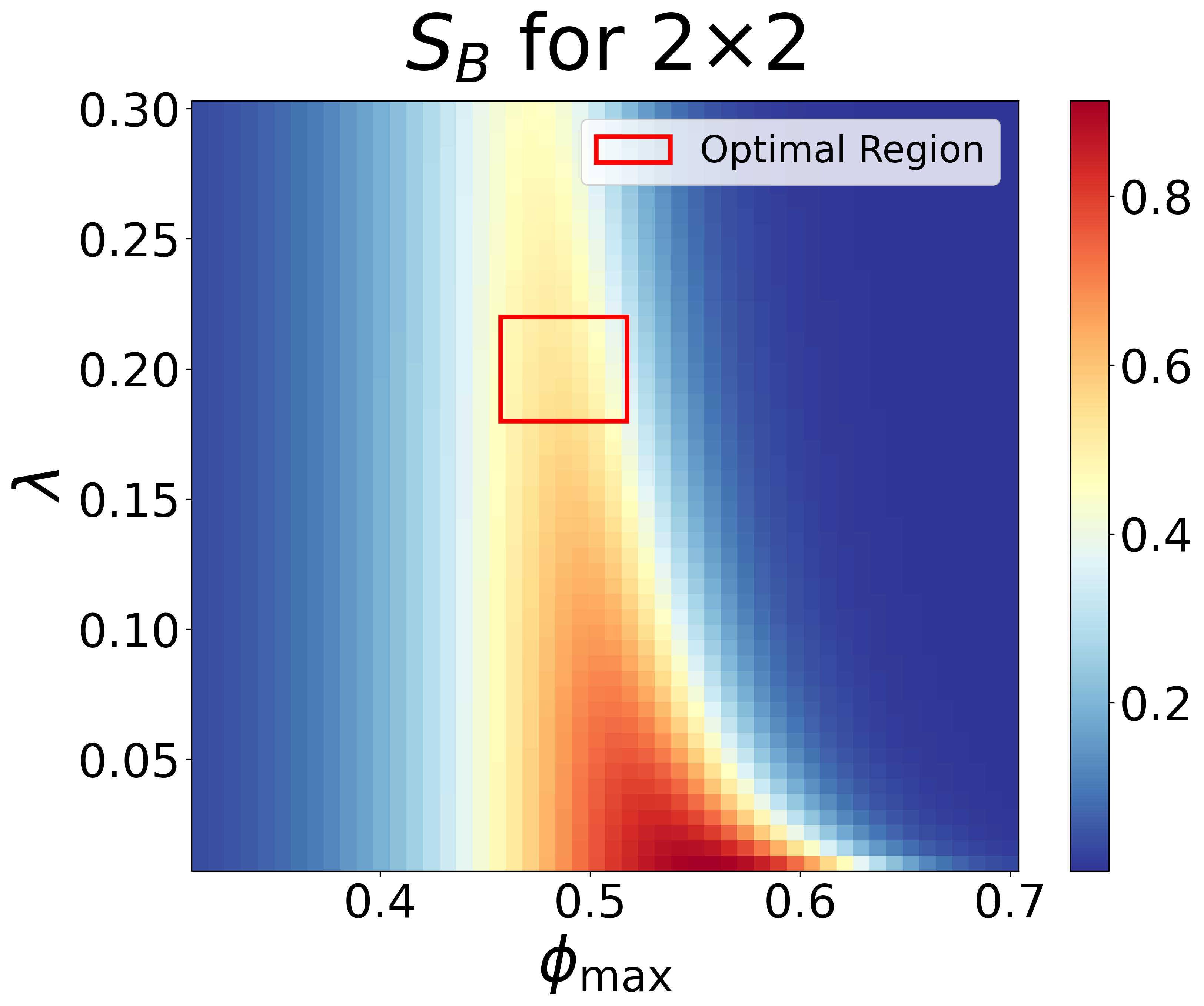}
\end{subfigure}\hfill
\begin{subfigure}{0.48\textwidth}
\centering
\includegraphics[width=\linewidth]{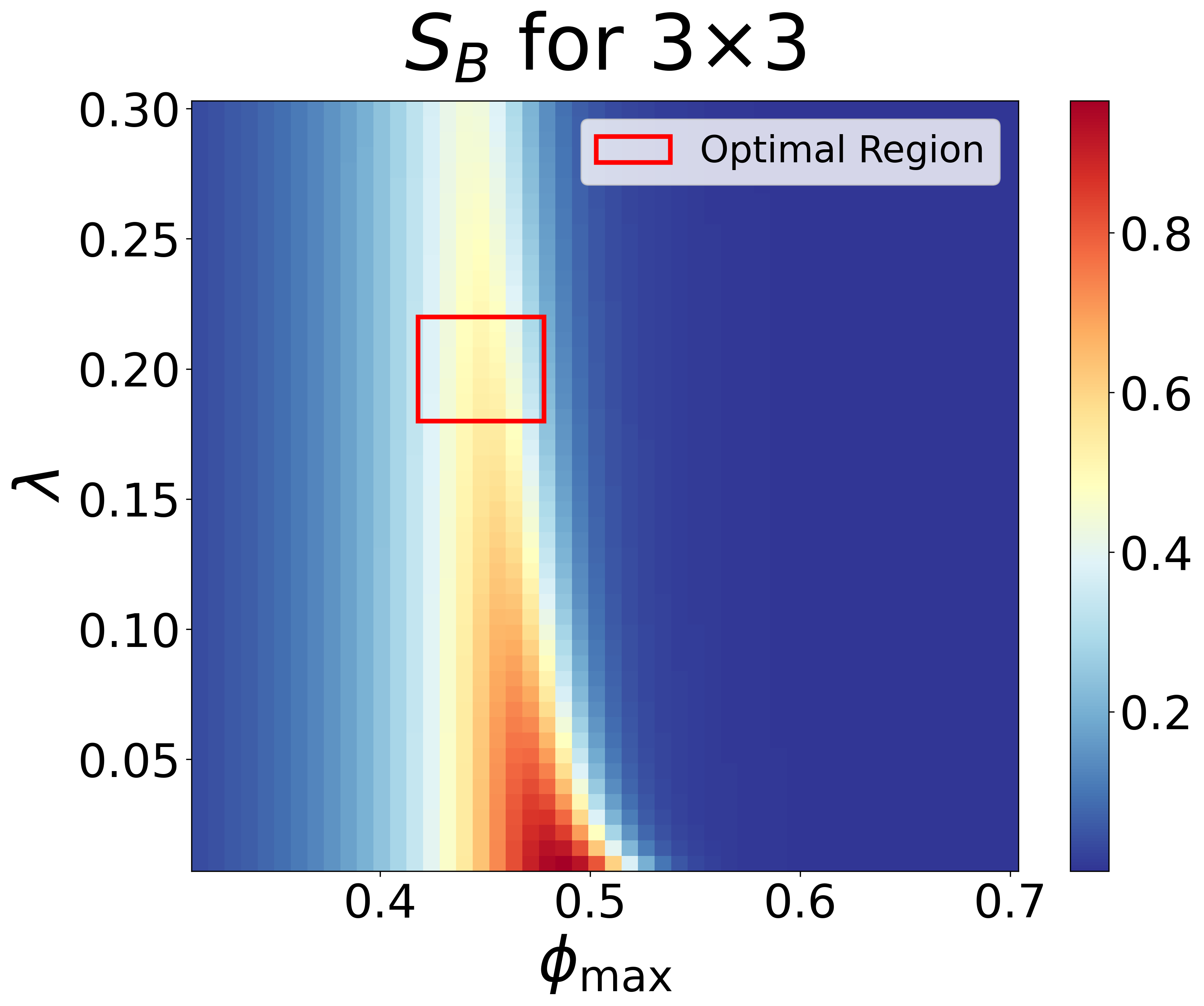}
\end{subfigure}

\caption{The energy gap $\Delta E$ and entanglement entropy $S_B$ of the ground state obtained with direct diagonalization.}
\label{fig:heatmap}
\end{figure}

To ensure high fidelity in the LLM-discovered quantum circuits, we employ a loss function that simultaneously minimizes both energy and infidelity. This approach is only feasible for the small $3\times3$ lattice we utilize to search for the ansatz. The loss function is given by
\begin{align}
    L \;=\; &10[\delta E]_+ \;+\; \bigl(80[0.95 - F]_+\bigr)^2 \;+\; 0.8[D-16]_+ \;+\; 0.3[N_{\texttt{CX}}-40]_+ \;+\; \bigl(1.5[N_{P}-4]_+\bigr)^2 \;+\; S_{\mathrm{stab}}(\sigma) \nn\\
    &+ \Big(\min\{(50[0.99-F_2]_+)^2,\,45\} + \min\{(150[0.94-F_4]_+)^2,\,45\}\Big),
\end{align}
where we define $[x]_+ \equiv \max(0, x)$ as a threshold of the penalty. Here $\delta E= E_{\mathrm{VQE}} - E_0$ is the energy deviation, $F$ the fidelity for the resulting state, $D$ the transpiled circuit depth, $N_{\texttt{CX}}$ the number of CNOT gates after transpilation, $N_P$ the total number of parameters of the circuit and $\sigma$ the standard deviation of the energy across 60 VQE runs with different initializations. The quantities $F_n$ denote the extrapolated fidelities on the $n\times n$ system. The first two terms penalize energy deviation and fidelity deficiency, respectively. The next three terms impose mild penalties on transpiled depth, the number of CNOT gates, and the number of parameter count. The stability term 
$$S_{\mathrm{stab}}(\sigma)=
\begin{cases}
5, & \sigma > 0.5,\\[2pt]
2, & 0.2 < \sigma \le 0.5,\\[2pt]
0, & \text{otherwise},
\end{cases}$$
favors circuits whose VQE optimization is stable (i.e., exhibits low variance across runs). We have included weighting coefficients to balance the contributions of different terms to the loss function. Finally, for extrapolation, we initialize the VQE on the $n\times n$ system with the parameter set optimized on $3\times 3$, perform a single VQE run, and then compute the $n\times n$ fidelity at the resulting parameters. The last penalty term applies to the fidelities obtained on $2\times2$ and $4\times 4$ under this protocol, promoting extrapolation across system sizes.

\subsection{C. State preparation on a quantum hardware}
In the experiment, we selected five systems of different sizes, corresponding to $n = 10, 12, 15, 20$, respectively. We consider the initial circuit's noise strength to be $1$, and vary the circuit's noise strength by replacing a single CNOT gate with three or five CNOT gates. For a system of size $n$, we define noise strength as ${\cal N}_s = N_\text{CNOT}/(2n-2)$,  where $N_\text{CNOT}$ is the number of CNOT gates in the circuit and $2n-2$ is the minimum CNOT gates in the state preparation circuit. The noise strengths $\mathcal{N}_s$ considered for different $n$ can be seen from Fig. \ref{fig:qiskit-fitting}.

For ${\cal N}_s = 1$, the circuit corresponds to the original circuit. For noise strengths of 3, each CNOT gate in the original circuit is replaced by 3 CNOT gates, respectively. For the cases of ${\cal N}_s = 1,3$, we only performed a single run for each. For the other noise strength cases with $\mathcal{N}_s < 3$, we randomly select the number of CNOT gates in the circuit and replace each selected CNOT gate with three CNOT gates to match the corresponding $\mathcal{N}_s$. To reduce statistical error, this process is repeated five times to generate five circuits, and the final result is obtained by averaging the outcomes of these five circuits.

For the original data, we use the jackknife method to compute the energy and standard error for each circuit. When the noise strength is not 1 or 3, we also apply the jackknife method to calculate the mean and standard error when averaging the results of the five circuits. Then, we fit the noise strength and energy to obtain the energy in the zero-noise limit. We consider three fits: a three-parameter exponential fit $E(\mathcal{N}_s) = a + b\,\exp(-c\,\mathcal{N}_s)$, a three-parameter quadratic fit $E(\mathcal{N}_s) = a + b\,\mathcal{N}_s + c~\mathcal{N}^2_s$, and a two-parameter linear fit $E(\mathcal{N}_s) = a + b\,\mathcal{N}_s$. Fig. \ref{fig:qiskit-fitting} shows the fitting results for different values of $n$.

\begin{figure}
    \includegraphics[width=0.49\textwidth]{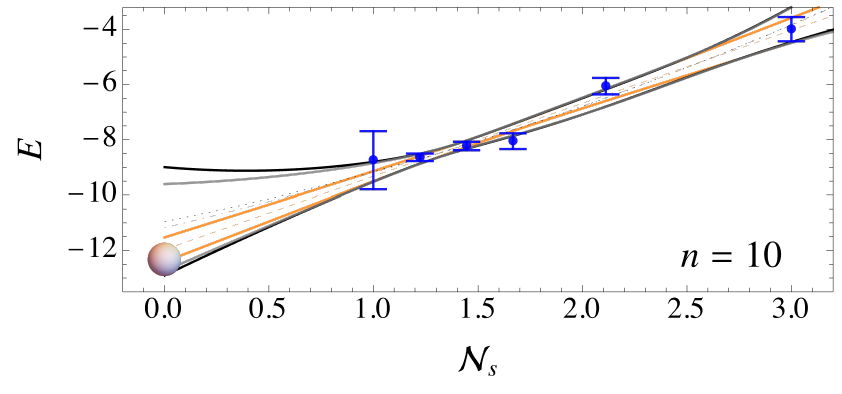}
    \includegraphics[width=0.49\textwidth]{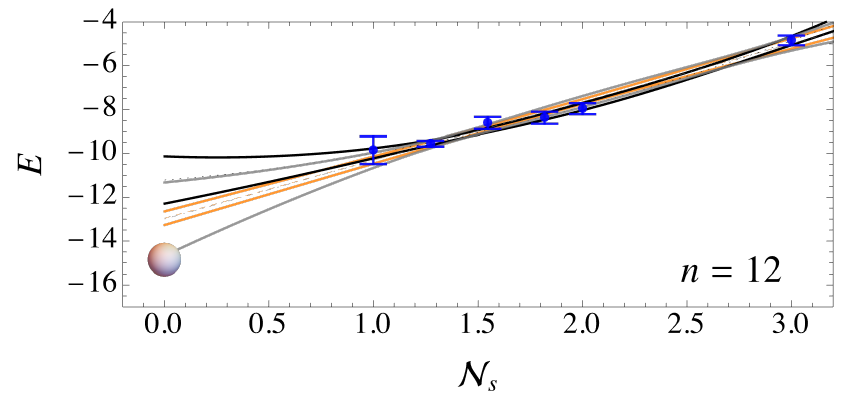}
    \includegraphics[width=0.49\textwidth]{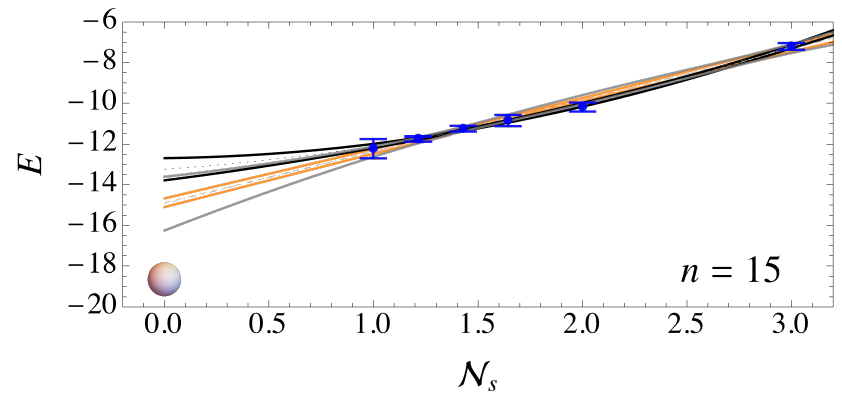}
    \includegraphics[width=0.49\textwidth]{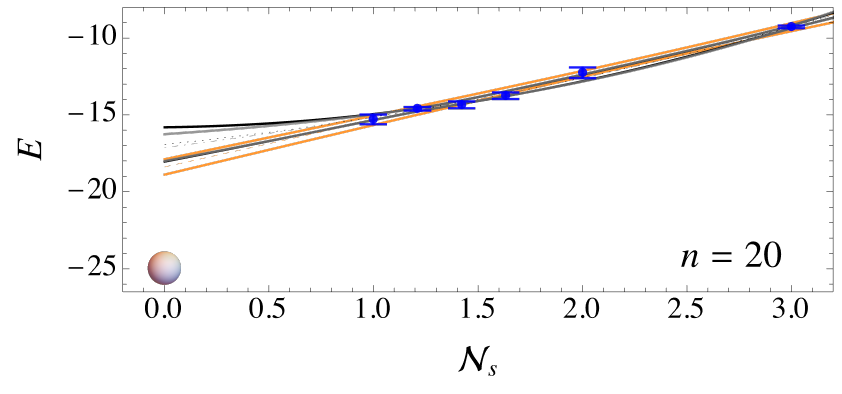}
  \caption{Fitting results for different system size $n$, with light gray dot-dashed lines representing the exponential fit, black dashed lines for the quadratic fit, and the orange dashed line for the linear fit. The bands, bounded by solid lines, indicate the $68\%$ confidence level (C.L.) uncertainties of the fits. The energies of the states prepared on the \texttt{Zuchongzhi} quantum chip at different values of $\mathcal{N}_s$ are shown as blue circles, with statistical error bars. The white circle at $\mathcal{N}_s$ indicates the energy of the state prepared by a noiseless quantum simulator.}
  \label{fig:qiskit-fitting}
\end{figure}

\end{widetext}

\end{document}